\documentclass[12pt]{article}
\usepackage{epsfig,a4}
\usepackage{color}
\def\wt{\widetilde}
\begin{document}
\title{The Bonn nuclear quark model revisited}
\author{
Constan\c{c}a Provid\^encia, Jo\~ao da Provid\^encia, Fl\'avio Cordeiro\\
{\it \small Departamento de F\'\i sica, Universidade de Coimbra,}\\
{\it \small  P-3004-516  Coimbra, Portugal}\\
Masatoshi Yamamura\\
{\it \small Department of Pure and Applied Physics, Faculty of
Engineering Science,}\\
{\it \small Kansai University, Suita 564-8680,  \small Japan}\\
Yasuhiko Tsue, Seiya Nishiyama\\{\it\small Physics Division, Faculty
of Science, Kochi University,}\\{\it\small Kochi 780-8520, Japan} }

\def\a{\alpha}
\def\b{\beta}

\maketitle \abstract{We present the exact solutions to the equations
of the lowest energy states of the colored and color-symmetric
sectors of the Bonn quark model, which is  $SU(3)$ symmetric and is
defined in terms of an effective pairing force with $su(4)$
algebraic structure. We show that the groundstate of the model is
not color symmetrical except for a narrow interval in the range of
possible quark numbers. We also study the performance of the Glauber
coherent state,  as well as of superconducting states of the BCS
type, with respect to the description, not only of the absolute
(colored) groundstate, but also of the minimum energy state of the
color-symmetrical sector, finding that it is remarkably good. We use
the model to discuss, in a schematic context, some controversial
aspects of the conventional treatment of color superconductivity.}
\section{Introduction}

The so-called Bonn model, which is here reinvestigated, has been
proposed by {Petry} et al. \cite{petry}  some years ago for the
description of the nucleus as a system of interacting quarks, which,
according to the current belief, is based on the QCD field theory.
The model is inspired by the famous seniority model of nuclear
physics that explains the superconducting features of nuclear
structure, namely, a characteristic gap in some spectra associated
with a pairing force (cf. \cite{ring}, p. 121). An important
ingredient in the Bonn model is an attractive pairing force, acting
between quarks of different colors, that suppresses physically
undesirable degeneracies of the quark system. Although the model is
too schematic to be realistic, it qualitatively accounts for some
features of nuclear physics. This model was originally aimed at
explaining the formation of color neutral triplets, that is,
clustering of quarks into nucleons. One of the reasons why it is not
realistic is that it contains a colored sector, to which the
groundstate belongs. Indeed, the existence of the colored sector is
a physically undesirable artifact of the model, since, according to
QCD, there are no colored states. Nevertheless, the model brings
color symmetrical states rather close to the groundstate. This is
quite remarkable since the involved interaction is a two-body force,
which is naturally associated with two-body correlations, but not
with three-body-correlations.

The model is particularly interesting because it is exactly soluble,
so that it is appropriate for testing certain classes of
approximation techniques. In this note we explore this aspect, by
studying the performance of the Glauber coherent state (GCS) and of
color superconducting states of the BCS type . It is found, quite
remarkably, that the GCS and the BCS states provide good
descriptions, not only of the absolute (colored) groundstate, but
also of the minimum energy state of the color-symmetrical sector. We
emphasize, however, that in order to properly describe the
physically important color-symmetrical sector, a modified BCS
approach, based on the generalized Bogoliubov transformation, which
treats the three colors on the same footing \cite{bohr}, is needed.

A BCS color superconducting state $|\Phi_{BCS}\rangle$ is
characterized by an order parameter $\wt\Delta$ which is a triplet
belonging to the $\bar 3$ representation of $SU(3)$, i.e.
$\wt\Delta=(\Delta_1,\Delta_2,\Delta_3)$. The starting point for the
color neutral version of the BCS approach, considered in Ref.
\cite{bohr}, is the assumption $\Delta_1=\Delta_2=\Delta_3=\Delta$,
but, by a convenient color rotation, this may be turned into
$\Delta_1=\sqrt{3}\Delta$, $\Delta_2=\Delta_3=0$, which is the
starting assumption for the standard approach  considered in Refs.
\cite{alford,alford1,iida}. Clearly, the groundstate energy remains
unchanged. These two assumptions are equivalent if one imposes the
restrictive condition
\begin{equation}\langle\Phi_{BCS}|S_{\Lambda_k}|\Phi_{BCS}\rangle=0,
\qquad k=1,\cdots,8, \label{csc1}\end{equation} where
$S_{\Lambda_k}$ are the $SU(3)$ generators associated with the Gell
Mann matrices ${\Lambda_k}$. On the other hand, for the standard BCS
approach, either these relations are ignored altogether or, at most,
the condition
\begin{equation}\langle\Phi_{BCS}|S_{\Lambda_3}|\Phi_{BCS}\rangle=
\langle\Phi_{BCS}|S_{\Lambda_8}|\Phi_{BCS}\rangle=0\label{csc2},
\end{equation} is imposed through color
dependent chemical potentials \cite{iida}. Since the condition
(\ref{csc2}) is not stable under color rotations, this is not
enough, if our goal is the description of the color-symmetric
states. In effective models of the NJL type, an un-physical colored
sector coexists with the physically relevant color-symmetric sector,
whose states are color singlets. As shown in Ref. \cite{bohr}, the
colored sectors of the Bonn model, or of the NJL model, are
described when the restrictive conditions (\ref{csc1}) are
altogether disregarded. On the other hand, the novel approach
presented in Ref. \cite{bohr} allows for the description of the
physically relevant colorless sector.

It is well known that the dynamics of many-fermion systems may be
described in terms of bosons. In the collective model of Bohr and
Mottelson (cf. \cite{ring}, p. 9), bosons were introduced through
the quantization of the oscillations of a liquid drop to describe
excitations of nuclei. In the theory of plasma oscillations, excited
states of the electron gas are described by the so called random
phase approximation which is nothing else than particle-hole pairs
being approximated by bosons. The  physical interpretation of the
Schwinger bosons in the case of the Bonn model is an interesting
problem which remains open.

We will present the exact solutions to the equations of the lowest
energy states of the colored and color-symmetrical sectors of the of
the Bonn quark model, and will show that the groundstate of the
model is not color-symmetrical except for a narrow interval in the
range of possible quark numbers. We will next study the performance
of the Glauber coherent state with respect to the description, not
only of the absolute (colored) groundstate, but also of the minimum
energy state of the color-symmetrical sector. The color-symmetric
sector of the Bonn model, characterized by its color singlet-ness,
will receive a special attention. In particular, we will present its
description, in terms of coherent states, such as color neutral
Glauber coherent states, or superconducting states of the
generalized BCS type introduced in \cite{bohr}.

\section{The model}

We wish to investigate the schematic nuclear model which has been
proposed by H.R. Petry  {\it et al.} \cite{petry}, and further
discussed by Pittel {\it et al.} \cite{pittel}, among others. Its
Hamiltonian reads
\begin{equation}\label{1}H=G\sum_{j=1}^3S^iS_i,
\end{equation} where
\begin{equation}S^1=S_1^\dagger=\sum_m c^\dagger_{2m}c^\dagger_{3\tilde m} ,\quad S^2=S_2^\dagger=\sum_m
c^*_{3m}c^\dagger_{1\tilde m}, \quad
S^3=S_3^\dagger=\sum_mc^\dagger_{1m}c^\dagger_{2\tilde m}
.\label{2}\end{equation}
 Here,
$c_{i m}^\dagger$ are fermion creation operators, the indexes $i,m$
denoting the color and the remaining single particle quantum
numbers. The quantum number $\tilde m$ is obtained from $m$ by time
reversal, so that $\tilde{\tilde m}=m$, or, more precisely, in a
standard notation,
$$c^\dagger_{i\tilde m}=(-1)^{j_s-m}c^\dagger_{i,~-m}.$$
 Clearly,
the operators $S^i,S_i$ generate an $su(4)$ algebra. Indeed we have
\begin{eqnarray}
 [ S^i, S^j]=0,\quad  [ S^i, S_j]=
S^j_i=-\sum_mc^\dagger_{jm}c_{im}+\delta_{ij}(\sum_{km}c^\dagger_{km}c_{km}-2\Omega),
\label{3}\end{eqnarray} where $2\Omega$ denotes the level
degeneracy, for fixed color, that is, the total number of single
particle states beyond color. Moreover, the following relations,
which close the $su(4)$ algebra, are satisfied.
\begin{eqnarray}
[ S^j_i, S^k]=\delta_{ij} S^k+\delta_{jk} S^i, \quad [ S^j_i,
S^k_l]=\delta_{jl} S^k_i-\delta_{ik} S^j_l. \label{4}\end{eqnarray}

\section{\bf Schwinger realization}

We use the  Schwinger realization  of $su(4)$ in terms of 8 bosons,
$a_1,a_2,a_3,a,b_1,b_2,b_3,b,$  
which has been proposed by Yamamura {\it et al.} \cite{yamamura},
and reads
\begin{eqnarray*}&&S^i=a^\dagger_ib+a^\dagger b_i,\quad S_i=b^\dagger a_i+b_i^\dagger a,\\&&
S_j^i=[S^j,S_i]=a_j^\dagger a_i-b_i^\dagger
b_j+\delta_{ij}(a^\dagger a-b^\dagger b).
\end{eqnarray*}
It may be easily checked that this realization preserves the algebra
of the operators $S^j,S_i,S^j_i$, that is, the commutation relations
in (\ref{3}), (\ref{4}) are satisfied.


 The transformed Hamiltonian reads
\begin{eqnarray*}H&=&G\sum_{i=1}^3(a_i^\dagger b^\dagger a_ib+a^\dagger
b_i^\dagger ab_i+a_i^\dagger b_i^\dagger ab+a^\dagger b^\dagger
a_ib_i)\\&& +G\sum_{i=1}^3(a_i^\dagger a_i+a^\dagger a).
\end{eqnarray*}
This Hamiltonian admits the following constants of motion
\begin{eqnarray*}
&&L_i=b_i^\dagger b_i-a_i^\dagger a_i,\quad i=1,2,3,\quad
L=b^\dagger b-a^\dagger a,\\&& Q_a=\left(\sum_{i=1}^3a_i^\dagger
a_i\right)+b^\dagger b, \quad Q_b=\left(\sum_{i=1}^3b_i^\dagger
b_i\right)+a^\dagger a .\end{eqnarray*} The constant of motion
$K=-\sum_{i=1}^3 (L_i+L)$ is related to the number $N$ of quarks.
From (\ref{3}), we have
\begin{eqnarray}K=2N-6\Omega\label{K}.\end{eqnarray} Moreover,
$Q_a+Q_b$ commutes with all the generators of the algebra and so its
eigenvalues characterize the physically relevant irreducible
representations, being then
\begin{eqnarray}S=Q_a+Q_b=2\Omega.\label{S}\end{eqnarray} We discuss now the
construction of the lowest weight state. We will not consider states
containing broken pairs. This means that our discussion is
restricted to states $|\Phi\rangle$ such that the numbers of
occupied single particle levels of the types $|i,m\rangle$ and
$|j,\tilde m\rangle$ are the same, for any $i,j$ not necessarily
different. Of course it is important to discuss also the case of
broken pairs. This point will be considered elsewhere. The minimum
weight state of the symmetric representation reads
\begin{eqnarray*}&&|\Omega\rangle={b^\dagger}^{2\Omega}|0\rangle,\end{eqnarray*} and
satisfies
\begin{eqnarray*} &&
S_i|\Omega\rangle =S_i^j|\Omega\rangle=0,\quad S_i^i
|\Omega\rangle=-2\Omega|\Omega\rangle,\quad
(S^{i})^{2\Omega}|\Omega\rangle=(2\Omega)!\,{a^\dagger_i}^{2\Omega}|0\rangle.
\end{eqnarray*} The state $|\Omega\rangle$ is the boson image of the Fermion
vacuum. In order to discuss asymmetric representations, we also
consider the state:
\begin{eqnarray*}&&|\Omega,\Omega'_3\rangle=b^{\dagger~2(\Omega-\Omega'_3)}b_3^{\dagger
~2\Omega'_3}|0\rangle.\end{eqnarray*}It satisfies:\begin{eqnarray*}
&& S_i|\Omega,\Omega'_3\rangle
=S_1^2|\Omega,\Omega'_3\rangle=S_2^1|\Omega,\Omega'_3\rangle=S_1^3|\Omega,
\Omega'_3\rangle=S_2^3|\Omega,\Omega'_3\rangle=0,\;i=1,2,3,\\
&&S_1^1 |\Omega,\Omega'_3\rangle=S_2^2
|\Omega,\Omega'_3\rangle=-2(\Omega-\Omega'_3)|\Omega,\Omega'_3\rangle,\quad
S_3^3 |\Omega,\Omega'_3\rangle= -2\Omega|\Omega,\Omega'_3\rangle,\\
&&(S^{i})^{2(\Omega-\Omega')}|\Omega,\Omega'_3\rangle=
(2\Omega-2\Omega')!\,a^{\dagger~2(\Omega-\Omega'_3)}_ib_3^{\dagger~2\Omega'_3}|0\rangle,\,i=1,2,\\&&
(S^{3})^{2\Omega}|\Omega,\Omega'_3\rangle=(2\Omega)!\,
{a_3^\dagger}^{2(\Omega-\Omega'_3)}{a^\dagger}^{2\Omega'_3}|0\rangle.
\end{eqnarray*}
The state $|\Omega,\Omega'_3\rangle$ is the boson image of the
fermionic state
$\prod_{m=1}^{\Omega'_3}c^\dagger_{3m}c^\dagger_{3\tilde
m}|0_F\rangle$. The generators $S_i^j$ commute with $H$. They
generate an $su(3)$ sub-algebra of $su(4)$. Thus, $H$ has $su(3)$
symmetry. It may be easily shown that the states
$(S^{i})^p|\Omega\rangle$ (containing $2p$ quarks) and
$(S^{3})^p|\Omega,\Omega'_3\rangle_{\Omega'_3=\Omega}$ (containing
$2p+2\Omega$ quarks) are eigenstates of $H$,
$$H(S^i)^p|\Omega\rangle=Gp(2\Omega+1-p)(S^i)^p|\Omega\rangle,
$$
$$H(S^3)^p|\Omega,\Omega'_3\rangle_{\Omega'_3=\Omega}
=Gp(2\Omega+3-p)(S^3)^p|\Omega,\Omega'_3\rangle_{\Omega'_3=\Omega}.
$$
For $0<\Omega'_3<\Omega$ we find
$${\langle p,\Omega,\Omega'_3|H|p,\Omega,\Omega'_3\rangle\over
\langle p,\Omega,\Omega'_3|p,\Omega,\Omega'_3\rangle}=
Gp\left(2\Omega+1+2{\Omega'_3\over\Omega}-p\right),
$$ where
$|p,\Omega,\Omega'_3\rangle=(S^{3})^p|\Omega,\Omega'_3\rangle.$ This
state contains $N=2(p+\Omega'_3)$ quarks. For fixed $\Omega'_3$,
this expectation value, as a function of $N$, is a parabola. The
envelope of these parabolas is the parabolic arc
$G{(N+\Omega+2\Omega^2)^2\over 4\Omega(2+\Omega)} $. Therefore, for
$2\Omega\leq N\leq4\Omega$, the groundstate energy is, according to
the Born approximation, $G{(N+\Omega+2\Omega^2)^2\over
4\Omega(2+\Omega)}$ $\approx G(\Omega^2-\Omega+N)$. For
$N\leq2\Omega,$ the groundstate energy is $GN(4\Omega+2-N)/4.$ For
$4\Omega\leq N\leq6\Omega,$ the groundstate energy is
$G(N-2\Omega)(6\Omega+6-N)/4.$ The Bonn model was originally devised
as a model for the formation of triples (clustering of quarks into
nucleons). However, the tendency for the formation of triplets is
not dominant. If it were, the groundstate would have $SU(3)$
symmetry, which is not the case. The investigation of the
groundstate energy beyond the Born approximation will be carried out
in Sections 5 and 6.

\section{\ Glauber coherent state}

It is believed that color superconductivity plays an important role
in high density matter, such as may be found in the core of neutron
stars \cite{alford,alford1,iida}. When the BCS theory is applied to
the superconducting phase of quark matter, it is found that the BCS
state violates color charge neutrality \cite{iida}. This undesirable
feature is attributed to the BCS theory itself. However, in the
model (\ref{1}) which is used here, or in models of the NJL type
\cite{iida}, color symmetry breaking states, which are un-physical,
coexist with the physically relevant states, which are color
symmetric. In \cite{bohr}, it was shown that superconducting BCS
states describe well the lowest energy colored and color-symmetric
states of the Bonn model. There, it was also shown that the
description of the physically relevant color-symmetric sector
requires a modified version of the conventional BCS approach based
on the generalized Bogoliubov transformation. Glauber coherent
states have some analogies with BCS states. So, it may be
interesting to investigate their performance. The GCS may be written
$$|\Phi\rangle=|\alpha_3,\alpha,\beta_3 ,\beta\rangle=\exp(\alpha_3\, a_3^\dagger+\alpha
a^\dagger\, +\beta_3\, b_3^\dagger+\beta\, b^\dagger)|0\rangle.
$$
The Hamiltoninan expectation value reads
\begin{eqnarray*}{\cal
E}={\langle\Phi|H|\Phi\rangle\over\langle\Phi|\Phi\rangle}&=&G(\a_3^*
\b^*\a_3\b+\a^*\b_3^*\a\b_3+\a_3^*\b_3^*\a\b +\a^*\b^*\a_3\b_3)\\&
+&G(\a_3^*\a_3+3\a^*\a) . 
\end{eqnarray*} The expectation values of the constants of motion read
\begin{eqnarray*}&&{\cal Q}_a=\langle\Phi|Q_a|\Phi\rangle=
\a_3^*\a_3+\b^*\b=2(\Omega-\Omega'),\\& & {\cal
Q}_b=\langle\Phi|Q_b|\Phi\rangle=\b_3^*\b_3+\a^*\a=2\Omega'\\&&
{\cal K}=\langle\Phi|K|\Phi\rangle=\a_3^*\a_3-\b_3^*\b_3+3(\a^*\a-\b^*\b).\end{eqnarray*} 
\begin{figure}[ht]
\centering
\includegraphics[width=0.6\textwidth, height=0.5\textwidth]{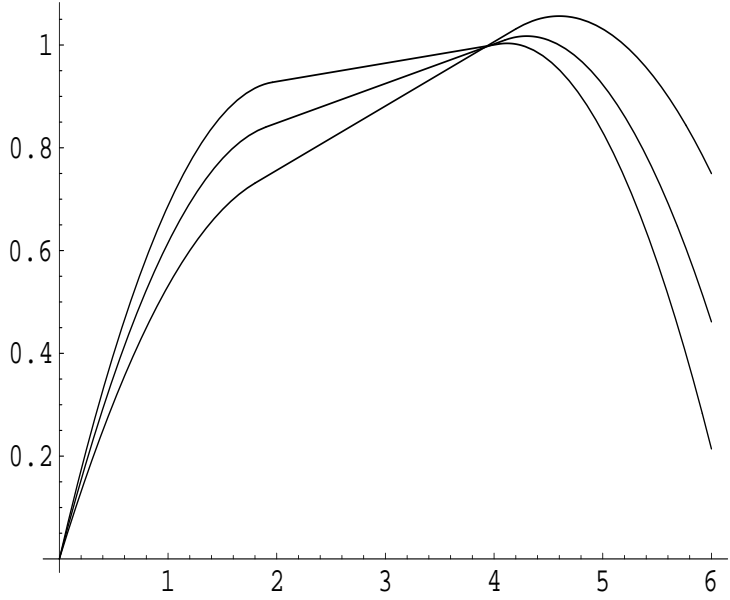}
\caption{  Groundstate energy, divided by $G\Omega(\Omega+3)$, vs.
$N/\Omega$, according to the Glauber coherent state, for $\Omega=5$
(lower curve for $N/\Omega\leq4$), $\Omega=10$ (middle curve for
$N/\Omega\leq4$), and  $\Omega=20$ (upper curve for $N/\Omega\leq4$)
. It almost coincides with the ``exact" result, which will be
presented later.} \label{BSfig1}
\end{figure}

\noindent We may chose $\arg(\a_3)+\arg(\b_3)-\arg(\a)-\arg(\b)=0.$
Then, we are led to minimize
\begin{eqnarray*}{\cal E}&=G(|\a_3|^2 |\b|^2 +|\a|^2 |\b_3|^2 +2 |\a_3| |\b|
|\b_3||\a| + |\a_3|^2+3|\a|^2) ,
\end{eqnarray*}
 with respect to $|\a_3|,|\b_3|,|\a|,|\b|,$ under the constraints
\begin{eqnarray*}&2\Omega=|\alpha_3|^2+|\beta_3|^2+|\alpha|^2+|\beta|^2,\cr&
 2N-6\Omega=|\alpha_3|^2-|\beta_3|^2+3(|\alpha|^2-|\beta|^2). 
 \end{eqnarray*}
\subsection{Computational details}

If $N\leq2\Omega-1$, the minimum energy occurs for
$|\beta_3|=|\alpha|=0 $. Then
$${\cal E}_0=G(|\alpha_3|^2|\beta|^2+|\alpha_3|^2),
$$
with $2\Omega=|\alpha_3|^2+|\beta|^2$,
$2N-6\Omega=|\alpha_3|^2-3|\beta|^2$, so that $ |\alpha_3|^2=N/2$,
$|\beta|^2=2\Omega-N/2$, and
$${\cal E}_0={G\over4}N(4\Omega-N+2).\eqno{(A)}
$$
If $4\Omega+1\leq N\leq6\Omega$, the minimum energy occurs for
$|\beta|=|\alpha_3|=0 $. Then
$${\cal E}_0=G(|\alpha|^2|\beta_3|^2+3|\alpha|^2),
$$
with $2\Omega=|\alpha|^2+|\beta_3|^2$,
$2N-6\Omega=3|\alpha|^2-|\beta_3|^2$, so that $
|\alpha|^2=N/2-\Omega$, $|\beta_3|^2=3\Omega-N/2$, and
$${\cal E}_0={G\over4}(N-2\Omega)(6\Omega+6-N).\eqno{(B)}
$$
If $2\Omega-1\leq N\leq4\Omega+1$, the minimum energy occurs for
$|\beta|\approx|\alpha_3|\neq0 $, $|\beta_3|\approx|\alpha|\neq0.$
This approximate result becomes exact in the limit of large
$\Omega$. Assuming $|\beta|=|\alpha_3| $, $|\beta_3|=|\alpha|$ we
may write
$${\cal E}_0=G((|\alpha|^2+|\beta|^2)^2+|\beta|^2+3|\alpha|^2),
$$
with $2\Omega=2|\alpha|^2+2|\beta|^2$,
$2N-6\Omega=2(|\alpha|^2-|\beta|^2)$, so that $
|\alpha|^2=N/2-\Omega$, $|\beta|^2=2\Omega-N/2$, and ${\cal
E}_0\approx G(\Omega^2-\Omega+N). $ If first order corrections are
consistently taken into account we find
$${\cal E}_0=G\left(\Omega^2-\Omega+N+{1\over4}\right).\eqno{(C)}
$$
Indeed, it may be seen that for fixed values of $\Omega$ and
$\Omega'=|\b_3|^2+|\a|^2,$ such that $0<\Omega'<\Omega$, the minimum
energy $\cal E$ as a function of $N$, lies on a similar parabola to
the one in eq. (A), but properly translated, so that it is tangent
to the line described by eq. (C).

In short, the groundstate energy reads
\begin{eqnarray*}&&{\cal E}_0= {G\over4}{N}(4\Omega-{N}+2) ,\quad\quad{\rm
if}\quad\quad 0\leq N\leq2\Omega-1;\\&&{\cal
E}_0={G}\left(\Omega^2-\Omega
+N+{1\over4}\right),\quad\quad\quad{\rm if}\quad\quad\quad
2\Omega-1\leq N\leq4\Omega+1;\\&&{\cal E}_0=
{G\over4}({N}-{2\Omega})({6\Omega}-{N}+6) ,\quad{\rm if}\quad
4\Omega+1\leq N,
\end{eqnarray*} which is very close to the exact result. This is quite
remarkable. Rather than being regarded as a model for the formation
of triples which are invariant under $SU(3)$, the Bonn model may be
viewed as a model for color superconductivity, similar to the
seniority model of nuclear physics, which is a model for nuclear
superconductivity.
\subsection{Glauber coherent states with color neutrality}
The condition for color neutrality reads
\begin{eqnarray*}&&\langle\Phi|S^1_{2}|\Phi\rangle=
\langle\Phi|S^2_{3}|\Phi\rangle=\langle\Phi|S^3_{1}|\Phi\rangle=0,\quad
\langle\Phi|S^1_{1}|\Phi\rangle=\langle\Phi|S^2_{2}|\Phi\rangle=
\langle\Phi|S^3_{3}|\Phi\rangle,\end{eqnarray*}
which implies
$$|\a_1|=|\b_1|,\;|\a_2|=|\b_2|,\;|\a_3|=|\b_3|.
$$
Then, we are led to minimize\emph{}
\begin{eqnarray*}{\cal E}=G\left(|\a_3|^2 (|\b|^2 +|\a|^2
+2 |\b| |\a|) + |\a_3|^2+3|\a|^2\right) ,
\end{eqnarray*}
 with respect to $|\a_3|,|\a|,|\b|,$ under the constraints
\begin{eqnarray*}2\Omega=2|\alpha_3|^2+|\alpha|^2+|\beta|^2,\quad
 2N-6\Omega=3(|\alpha|^2-|\beta|^2).
 \end{eqnarray*} This amounts to minimizing, with respect to $|a|$,
 the quantity
 \begin{eqnarray*}{\cal E}=G\left[2|\a|^2+{1\over3}N+2\left({1\over3}N-|\a|^2\right)
 \left(\Omega+|\a|^2-{1\over3}N+\sqrt{|\a|^2\left(2\Omega+|\a|^2-{2\over3}N\right)}\,\right)\right].
 \end{eqnarray*}
 The corresponding minimum is plotted, as a function of $N$, in Fig.
 3.
\begin{figure}[ht]
\centering
\includegraphics[width=0.6\textwidth, height=0.5\textwidth]{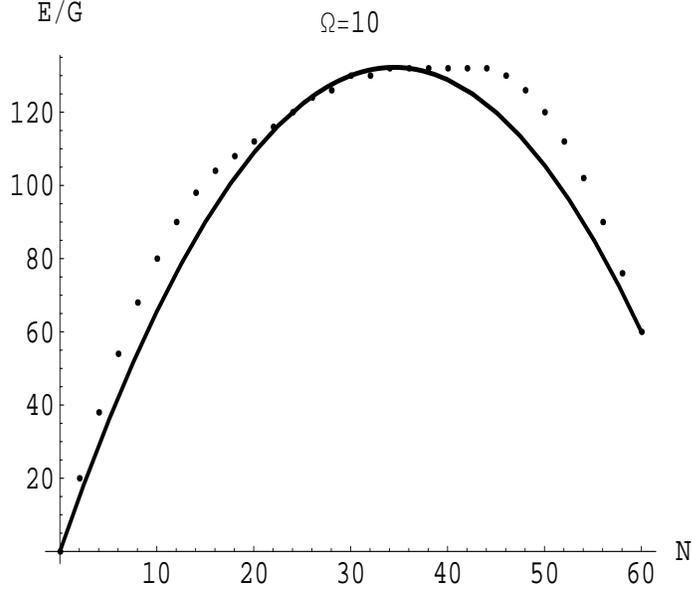}
 \caption{Exact results are shown. Absolute groundstate energy for  even particle numbers
(dots), and groundstate energy of the color neutral sector
(continuous line) vs $N$, for $\Omega=10$. The continuous line is
meaningful for values of $N$ multiples of 6. For $N=30$ the
groundstate is color symmetrical.} \label{BSfig2}
\end{figure}

\section{Color-symmetrical states -- exact treatment}
\bigskip
Consider the color symmetrical subspace spanned by the
non-normalized kets
\begin{eqnarray}|\Psi(m,\Lambda,\Gamma)\rangle=\left(\sum_i
a_i^\dagger b_i^\dagger\right)^m(a^\dagger
b^\dagger)^{\Lambda-m}(b^\dagger)^\Gamma|0\rangle,\label{star}
\end{eqnarray} which are clearly invariant under $SU(3)$.
This is an invariant subspace of $H$ of dimension $\Lambda+1$. From
(\ref{K}), (\ref{S}) we have
$2\Omega=2\Lambda+\Gamma,\;2N-6\Omega=-3\Gamma.$ Moreover,
\begin{eqnarray*}H|\Psi(m,\Lambda,\Gamma)\rangle
&=&Gm(m+2)|\Psi(m-1,\Lambda,\Gamma)\rangle\\&&
+G((2\Lambda+\Gamma-2m)m+(3\Lambda-2m))|\Psi(m,\Lambda,\Gamma)\rangle\\&&
+G(\Lambda-m)(\Lambda+\Gamma-m)|\Psi(m+1,\Lambda,\Gamma)\rangle,
\end{eqnarray*}
where $m=0,1,\cdots, \Lambda$. This provides a
$(\Lambda+1)\times(\Lambda+1) $ matrix representation of $H$ in the
color-symmetrical sector. Numerically, the eigenvalues of the matrix
so obtained may easily be determined. However, to obtain the
groundstate energy, no diagonalization is required. We remark that
the matrix which represents $G^{-1}H$ has non-negative entries and
that the sum of the entries in any row is the same as for any other
row. It is a well known result in matrix theory that this sum is the
largest eigenvalue. In the present case the sum reads
$m(m+2)+(\Lambda-m)(\Lambda+\Gamma-m)+(2\Lambda+\Gamma-2m)m+(3\Lambda-2m)=\Lambda(\Lambda+\Gamma+3),
$ independently of the value of $m$. It follows, therefore, that
$${\cal E}_0=G\Lambda(\Lambda+\Gamma+3)=G{N\over3}\left(2\Omega+3-{1\over3}N\right)$$
is the groundstate eigenvalue of the matrix representation of $H$,
in the color-symmetric subspace. It is the groundstate energy of the
color-symmetrical sector for $3\Omega\leq N\leq 6\Omega$. We remark
that the previous expression also holds for the sector $0\leq N\leq
3\Omega$, which may be similarly investigated by interchanging, in
(\ref{star}), $a^\dagger$ with $b^\dagger$.
\begin{figure}[ht]
\centering
\includegraphics[width=0.6\textwidth, height=0.5\textwidth]{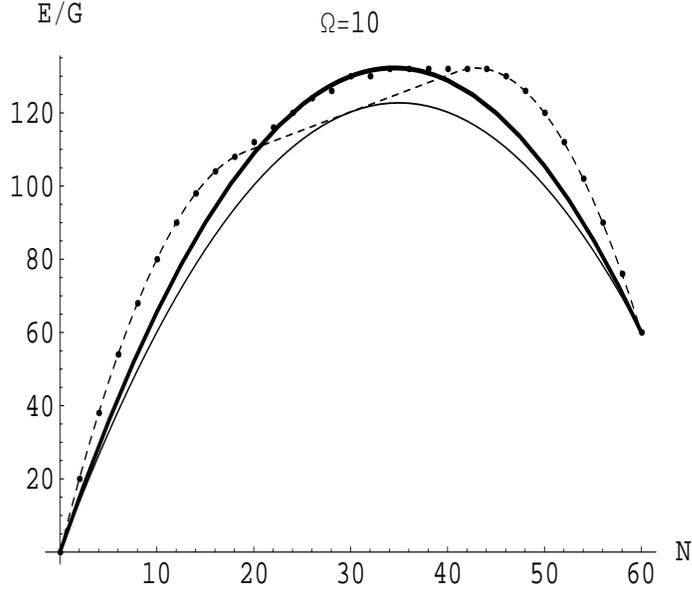}
\caption{Exact groundstate energy for even particle numbers: dots.
Glauber groundstate energy:
dashed line. Lowest energy in the color neutral sector: 
thick line, which is meaningful for values of $N$ multiples of 6
. Lowest energy in the Glauber color neutral state: thin line . For
$N\leq 2\Omega$ and $N\geq 4\Omega$, the exact result coincides with
the Glauber result for the groundstate. } \label{BSfig3}
\end{figure}
\begin{figure}[ht]
\centering
\includegraphics[width=0.6\textwidth, height=0.5\textwidth]{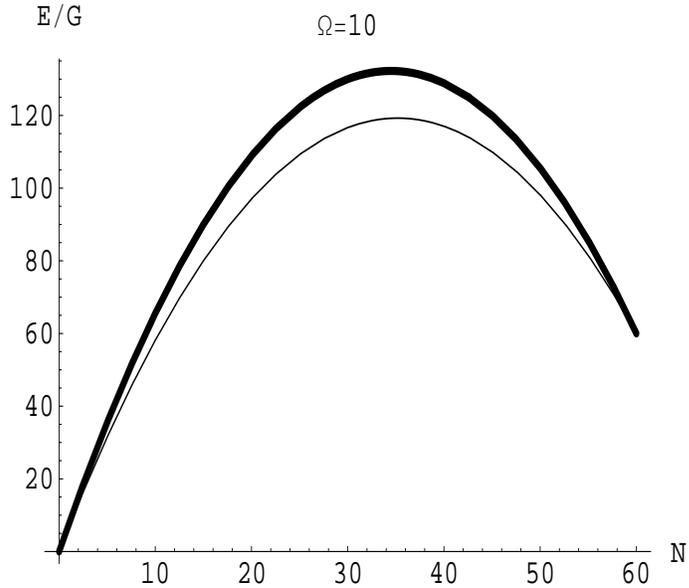}
\caption{Groundstate energy of the color symmetrical sector versus
the quark number, for $\Omega=10$. Thick line: exact result; thin
line: color symmetrical BCS estimate .} \label{fig1}
\end{figure}
\section{ Invariant sub-spaces of $H$}
\bigskip
The subspaces spanned by non-normalized kets of the any one of the
following forms \begin{eqnarray*}
|\Psi_{aa}(m,\Lambda,\Gamma,\Upsilon)\rangle=\left(\sum_i
a_i^\dagger b_i^\dagger\right)^m(a^\dagger
b^\dagger)^{\Lambda-m}(a^\dagger)^\Gamma(a_1^\dagger)^\Upsilon|0\rangle,
\end{eqnarray*}
$$|\Psi_{ab}(m,\Lambda,\Gamma,\Upsilon)\rangle=\left(\sum_i
a_i^\dagger b_i^\dagger\right)^m(a^\dagger
b^\dagger)^{\Lambda-m}(a^\dagger)^\Gamma(b^\dagger_1)^\Upsilon|0\rangle,
$$
$$|\Psi_{ba}(m,\Lambda,\Gamma,\Upsilon)\rangle=\left(\sum_i
a_i^\dagger b_i^*\right)^m(a^\dagger
b^\dagger)^{\Lambda-m}(b^\dagger)^\Gamma(a^\dagger_1)^\Upsilon|0\rangle,
$$
$$|\Psi_{bb}(m,\Lambda,\Gamma,\Upsilon)\rangle=\left(\sum_i
a_i^\dagger b_i^\dagger\right)^m(a^\dagger
b^\dagger)^{\Lambda-m}(b^\dagger)^\Gamma(b^\dagger_1)^\Upsilon|0\rangle,
$$
\noindent where $m=0,1,\cdots,\Lambda$, are also invariant subspaces
of $H$. Similar techniques to those used to obtain the groundstate
energy of the color-symmetrical sector may be used to obtain the
groundstate energy in each invariant subspace. For instance, we have
\begin{eqnarray*}&&G^{-1}H|\Psi_{aa}(m,\Lambda,\Gamma,\Upsilon)\rangle=
m(m+2+\Upsilon)|\Psi_{aa}(m-1,\Lambda,\Gamma,\Upsilon)\rangle\\&
&+[(\Lambda-m)(2m+\Upsilon)+m(\Gamma-2)+3\Lambda+3\Gamma+\Upsilon]|\Psi_{aa}(m,\Lambda,\Gamma,\Upsilon)\rangle\\&
&+(\Lambda-m+\Gamma)(\Lambda-m)|\Psi_{aa}(m+1,\Lambda,\Gamma,\Upsilon)\rangle.
\end{eqnarray*}  \noindent The sum of the above expansion
coefficients is independent of $m$ and reads
$$(\Lambda+1)\Upsilon+(\Lambda+\Gamma)(\Lambda+3).
$$
In the invariant subspace spanned by the kets
$|\Psi_{aa}(m,\Lambda,\Gamma,\Upsilon)\rangle$, the lowest energy is
therefore $G[(\Lambda+1)\Upsilon+(\Lambda+\Gamma)(\Lambda+3)], $ the
parameters $\Lambda,\Gamma,\Upsilon$, being restricted by the
relations $$2\Omega=2\Lambda+\Gamma+\Upsilon,\quad
2N-6\Omega=3\Gamma+\Upsilon.
$$ Combining the results obtained for each invariant subspace,
the  absolute groundstate energy is finally found. The results of
this procedure are shown in Fig. 2, for $\Omega=10$. In Fig. 3, the
results of the Glauber coherent state approach for $\Omega=10$, are
also presented, and compared with the exact results. The groundstate
is not color-symmetrical, except for $N=3\Omega$. Thus, in general,
the groundstate is not made up of color neutral triplets. We remark
that the lowest energy curve of the color neutral sector is
``tangent" to the absolute groundstate energy curve. Similarly, for
the Glauber coherent state, the lowest energy curve of the color
neutral sector is also tangent to the absolute groundstate energy
curve, in the corresponding approximation.

In Fig. 4 we illustrate the performance of the color-symmetric BCS
approach developed in \cite{bohr} for the description of the
color-symmetric sector of the model, being $${\cal
E}_{BCS}={GN\over9}\left(6\Omega-N+1+{4N\over3\Omega}\right),
$$
the color-symmetric groundstate energy in that approximation
\cite{bohr}.

\section{Conclusions}
Exact solutions are presented to the equations of the lowest energy
states of the colored and color-symmetrical sectors of the $SU(3)$
symmetric Bonn quark model, which is defined in terms of an
effective pairing force with $su(4)$ algebraic structure. It is
shown that the groundstate of the model belongs to the unphysical
color symmetry breaking sector, except for a narrow interval in the
range of admissible quark numbers. The exact treatment of the model
was carried out in the framework of the Schwinger realization of the
$su(4)$ algebra. The performances of the Glauber coherent state and
of the BCS theory with respect to the description, not only of the
absolute (colored) groundstate, but also of the minimum energy state
of the color-symmetrical sector, have been studied, it being found
that they are remarkably good. However, the description of the
color-symmetrical sector requires proper color-symmetric versions of
the Glauber coherent state and of the BCS theory.
\section*{Acknowledgments}
The present research was partially supported by projects
POCI/FP/81923/2007  and CERN/FP/ 83505/2008.


\begin{thebibliography}{999}
\bibitem{petry}  H.R. Petry, H. Hofstaedt, S. Merk, K. Bleuler, H. Bohr
and K.S. Narain, Phys. Lett. B159 (1985)363.
\bibitem{ring} P. Ring and P. Schuk, {\it The nuclear many-body
problem}, Springer-Vrlag, New-York, Heidelberg, Berlin, 1980.
\bibitem{pittel} S. Pittel, J. Engel, J.
Dukelsky and P. Ring, Phys. Lett. B247 (1990) 185.
\bibitem{bohr} H. Bohr and J. da Provid\^encia, J. Phys. A:
Math. Theor.  41 (2008) 405202;  H. Bohr and J. da Provid\^encia, J.
Phys. A: Math. Theor. 42 (2009) 089802, corrigendum.
\bibitem{alford} M. Alford, K. Rajagopal, S. Reddy and F. Wilczek,
Phys. Rev. D64 (2001) 074017; K. Rajagopal and F. Wilczek, Phys.
Rev. Lett. 86 (2001) 3492; M.G. Alford, Annu. Rev. Nucl. Part. Sci.
51 (2001) 131; M. Alford and S. Reddy, Phys. Rev. D67 (2003) 074024.
\bibitem{alford1} M.G. Alford, K. Rajagopal, T. Shaefer, A. Schmitt,
Rev. Mod. Phys. 80 (2008) 1455-1515.
\bibitem{iida}K. Iida and G. Baym, Phys.
Rev. D63 (2001) 074018; M. Alford and M. Rajagopal, J. High Energy
Phys. 06 (2002) 031; A.W. Steiner, S. Reddy and M. Prakash, Phys.
Rev. D66 (2002)094007; M. Huang, P. Zhuang and W. Chao, Phys. Rev.
D67 (2003) 065015.
\bibitem{yamamura}  M. Yamamura, T.
Kuriyama, A. Kunihiro, Prog. Theor. Phys. 104 (2000) 385.


\end{thebibliography}
\end{document}